\documentclass[article]{revtex4}%
\usepackage{amsmath}
\usepackage{graphicx}
\usepackage{epsfig}
\usepackage{dcolumn}
\usepackage{bm}
\usepackage{dcolumn}
\usepackage{amsfonts}
\usepackage{amssymb}%
\providecommand{\U}[1]{\protect\rule{.1in}{.1in}}
\begin{document}
\title{ On radiation reaction and the  Abraham-Lorentz-Dirac equation}
\author{A. Cabo Montes de Oca$^{*,**}$ and  Jorge Castineiras$^{*}$ \bigskip}
\affiliation{$^{*}$ Programa de Pos-Graduacao em F\'isica (PPGF) da Universidade Federal do
Par\'a (UFPA), Av. Augusto Correa, No. 01, Campus B\'asico do Guam\'a,
Bel\'em, Par\'a, Brasil \bigskip}
\affiliation{$^{**}$ Departamento de F\'isica Te\'orica, Instituto de Cibern\'etica
Matem\'atematica y F\'{\i}sica (ICIMAF), Calle E, No. 309, entre 13 y 15,
Vedado, La Habana, Cuba}

\begin{abstract}
\noindent It is underlined that the Lienard-Wiechert solutions indicate  that after the external
force  is instantly removed from a small charged particle,  the field in its close neighborhood becomes a  Lorentz boosted  Coulomb field.  It suggests that the force of the self-field on the particle should instantaneously  vanish after a sudden removal of the external force.  A  minimal modification of Abraham-Lorentz-Dirac  equation is searched seeking to implement this property.  A  term assuring this behaviour  is  added to the equation by maintaining  Lorentz covariance and vanishing scalar product with the four-velocity. The simple  Dirac´s constant force example does not show  runaway acceleration.

\bigskip

\noindent   A. Cabo E-Mail:  cabo@icimaf.cu,  \\
\noindent   J. Castineiras E-Mail: jcastin@ufpa.br

\end{abstract}
\maketitle



The correctness of the Abraham-Lorentz-Dirac (ALD) equations for the point
particles interacting with their own fields had been a recurrent issue in
searching for a consistent formulation of Classical Electrodynamics
\cite{Lorentz,abraham,poincare, schot,Dirac,
Teitelboim,Landau,Rohrlich,Yighjian,Spohn}. The formulation of the equations,
already from the start, presented theoretical difficulties which had been
discussed along decades in the extensive literature devoted to theme. The
presence of solutions with values growing without limit (runaway behavior) or
pre-accelerated motions in advance to the applied forces, were the basic
un-physical properties exhibited by the ALD equations \cite{Yighjian,Spohn}. \

A relevant advance in the analysis of these difficulties was done in Ref.
\cite{Yighjian}. In this work, it was argued that the ALD equations can't be
derived in a completely exact way from the integral equations. This happens in
the coupled system formed by the charged body having the electric charged
concentrated in its surface and interacting with its electromagnetic field,
when the force has a non analytic behavior. \ Moreover, the author was able
to derive a correction to the ALD equations which should be introduced in
order to consider suddenly changing forces. The change corresponded to
multiplying the reaction force of the field on the particle by a proper-time
dependent function $\ \eta(\tau)$ which vanish at the proper time value at
which the forces is started to be applied and attains a unit value in a small
time interval of the order of the time required by light to travel the radius
of the considered structured body studied in Ref. \cite{Yighjian}. This change
was argued to allow eliminating both: the runaway and the pre-accelerated
types of solutions. Thus, in general terms, it can be concluded that the ALD
equations are not exact consequences of the integral equations of motions for
the coupled system describing the structured particle investigated in Ref.
\cite{Yighjian}. It can be also remarked that, since the size $a$ of the
system considered in the work is arbitrary, the elimination of the un-physical
nature of the solutions is possible for whatever arbitrary small value of $a$
is chosen. This indicates that the solution might be valid also in the point
like limit of the particle structure.

\ Although, the described study constitute a relevant step in the
understanding of the problem, the wrote modification of the ALD equations
seems somewhat restricted by the particular process of applying a force
showing a non analytic proper time dependence given by a sudden jump from zero
to a finite value at a given instant. Henceforth, it makes sense to
investigate the possibility of formulating the point particle equations of
motion in alternative form exhibiting causality and absence of runaway solutions.

In this work we intend to advance a formulation of the equation of motion in a
form directly showing causality. The discussion, is suspected that could
implement the main physical elements leading to the elimination of the runaway
and pre-accelerated solutions which were discovered in Ref. \cite{Yighjian}.
The basic idea motivating the search is the following observation: when an applied  force is suddenly removed
from acting over  a point particle, the Lienard-Wiechert solutions for the self-field surrounding it, shows
that after an arbitrarily short time interval $\tau$ passed,  the field  solution within a sufficiently small neighborhood of the particle becomes  rigorously given by a  Lorentz boosted  Coulomb field. But, a Lorentz boosted Coulomb field is not expected to exert any force on the particle generating it. This known simple property strongly suggests,
that the force applied by the self-field on the particle should also instantaneously  vanish after the
sudden removal of the force.  Therefore, since the ALD equations are widely expected to be valid for continuously varying  forces, a minimal modifications of them is here searched in order to implement the instantaneously vanishing
 of the acceleration after a sudden removal of the applied force.

The proposal for the modified equation of motion is simple: to add a term of
the same order of the Schott and Larmor one being jointly determined by the
dynamics and by the external force. The dynamic part of the new terms is
simply linear in the four-acceleration and the element defined by the forces is a
Lorentz invariant function of the proper time. It is completely defined by an
assumed piecewise continuous external force as a function of the proper time.
The linear in the acceleration dependence of the added contribution assures
that its scalar product with the four velocity vanishes, which is a
consistency condition of the relativistic invariant ALD equations. In this
letter, in first place the modified equation is presented and the satisfaction
of the equations argued for solutions being linear motions in the intervals of
vanishing forces, and  given by motions solving the ALD equations in the
region for which the forces are finite. Next, the equations are solved for the
linear motion in which a constant force in the observer's frame is applied
during a finite time interval. This is one simple motion discussed by Dirac in
its classical article. For this case he encountered for the first time the
situation in which the acceleration (due to its continuity required by its
second derivative appearing in the equations), after a sudden elimination of
the applied force remained non vanishing and more drastically, it did no
decreased to a vanishing value continuously, but increased indefinitely. The
presented solution eliminates this problem because the new term adds a Dirac's
delta contribution in the proper time, which just cancels the also Delta like
terms appearing with the acceleration instantaneously jump to its zero value.
That is, the modification done simply allows the discontinuous transition
between the accelerated motion solution of the ALD equation (just existing
before the instant in which the force is eliminated) to the also solving the
ALD equation, uniform motion state. One point of interest to remark is that
the presented alternative equation seems to be, at least, coherent with the
picture of the electromagnetic propagation as described by the
Lienard-Wiechert potentials. These basic equations for the interaction of this
field with charged objects, indicate that when a particle
is instantly acquiring a state of uniform motion, at the exactly the same
instant, a wavefront start propagating from the point in which the particles
is. Inside the wavefront the field is the one associated to a moving charge
with constant velocity. Therefore, the self forces acting on the particle in
this interval should be zero. The modified equations, since they predict
vanishing of the acceleration after removing the forces, is at least
compatible with this property and solves the difficulty with the runaway
solution after the force suddenly vanishes \cite{berkeley}.
  The discussion presented also suggests a way for the discretization of a
general trajectory produced by a continuous in the proper time external force,
in a way which could probably eliminate the instabilities that currently beset
the numerical solution of the ADL equation. The idea is that by a  discretization
of the movement  in  infinitesimal  time intervals in which the forces periodically vanishes,
and modifying  the ALD equations in the described way, the
 developing of the numerical instabilities of the usual ALD equation  (being valid in the intervals in which  the force
 is not vanishing) could be eliminated.  These questions are expected to be  studied
 in extensions of the work to be considered elsewhere.

The letter proceeds as follows. Firstly the modified ALD equations and its general solution are presented. Next, the constant force case is discussed

Let us consider a general motion of a particle $P$ along a space-time
trajectory $C$ defined by a curve $x$($\tau)=(x_{0}^{0}(\tau),x^{1}%
(\tau)\mathbf{,}x^{2}(\tau),x^{3}(\tau))$ and parameterized by the proper time
$\tau.$ The metric signature to be used will be $(1,-1,-1,-1)$. \ The modified
ALD equations are proposed in the \ specific form%
\begin{align*}
f^{\mu}(\tau)\text{\ } &  \text{=}m\text{ }a^{\mu}(\tau)-\kappa\text{ }%
\frac{da^{\mu}(\tau)}{d\tau}-\kappa\text{ }a^{\nu}(\tau)a_{\nu}(\tau)\text{
}u^{\mu}(\tau)\text{ }+\kappa\text{ }a^{\mu}(\tau)\text{ }\sigma
(\tau),\text{\ }\\
u^{\mu}(\tau)\text{ } &  \text{=\ }\frac{dx^{\mu}(\tau)}{d\tau}\text{\ ,}\\
a^{\mu}(\tau)\text{ } &  \text{=\ }\frac{du^{\mu}(\tau)}{d\tau}\text{\ ,}%
\end{align*}
where the proper time function $\sigma(\tau)$ will be defined in what follows.
The equations satisfies the relativistic condition consisting in that the
scalar product of the four-velocity with them should vanish. This follows form
the form of the new added term, since
\[
\kappa\text{ }u^{\mu}(\tau)\text{ }a_{\mu}(\tau)\sigma(\tau)=0.
\]
The scalar function of the proper time $\sigma(\tau)$ will be chosen to be
determined by the force. Let us assume that the force is a piecewise
continuous function of the proper time $\tau$, and define the sequences of
proper times values $(s_{n},q_{n}),$ where $s_{n}$ is the set of increasing
with the integer value of $n,$ values of the proper time when the forces
discontinuously turns its values from zero to a finite one. \ The values of
$q_{n}$ \ indicate the instants at which the force is removed. \ An instant
before $q_{n}$ the force is finite and an instant after, it vanishes. Further,
in the pair $(s_{n},q_{n})$, the time $s_{n}$ is assumed to be one at which
the force is "connected" for afterwards being "disconnected" only at the
subsequent instant $q_{n}.$ \ A first step in the definition of $\sigma(\tau)$
is to define the function $\beta(\tau)$ in the form
\begin{align*}
\text{ }\beta_{\epsilon}(\tau) &  =\left\{
\begin{array}
[c]{cc}%
0 & \tau\in(q_{n-1},s_{n})\\
\frac{\tau-s_{n}}{\epsilon} & \tau\in(s_{n},s_{n}+\epsilon)\\
1 & \tau\in(s_{n}+\epsilon,q_{n}-\epsilon)\\
1-\frac{\tau-(q_{n}-\epsilon)}{\epsilon} & \tau\in(q_{n}-\epsilon,q_{n})\\
0 & \tau\in(q_{n},s_{n+1})
\end{array}
\right.  \\
\tau &  \in(q_{n-1},s_{n+1}),\text{ \ \ \ }\epsilon>0.
\end{align*}

Note for all value of $n$ the function becomes defined at all times. \ The
small $\epsilon$ parameter appearing is introduced in order that the
derivative over the proper time of the function leads to Dirac delta functions
with their support defined in the internal part of the intervals $(s_{n}%
,q_{n}).$ That means that the delta functions under integration will gives the
values of the limits at the boundaries of the function when taken from the
interior points of the intervals $(s_{n},q_{n})$.

The function $\sigma(\tau)$ is defined as the proper time derivative of
$\beta(\tau)$ when the limit $\epsilon\rightarrow0$%
\begin{align*}
\sigma(\tau)  &  =\text{\ }\lim_{\epsilon\rightarrow0}\frac{d\text{ }%
\beta_{\epsilon}(\tau)}{d\tau}\\
&  =\sum_{n=1}(\delta^{(+)}(\tau-s_{n})-\delta^{(-)}(\tau-q_{n})),
\end{align*}
and the plus and minus signs as superindices in $\delta^{(+)}(\tau-s_{n})$ and
$\delta^{(-)}(\tau-q_{n})$, indicate that the supports of the Delta functions
are at the interior of the intervals $(s_{n},q_{n})$. This completes the
definition of the modified ALD equations. \

\ Lets us present an explicit constructions of solutions of these equations.
For this purpose consider the following definitions. Assume that the
acceleration of the proposed solutions is only non vanishing in all intervals
$(s_{n},q_{n})$. That is
\[
a^{\mu}(\tau)=\sum_{n=1}\theta(\tau-s_{n})\theta^{(-)}(q_{n}-\tau)a_{n}^{\mu
}(\tau),
\]
where $a_{n}^{\mu}(\tau)$ is the acceleration of a solution of the ALD
equations in the considered interval, which satisfies initial conditions
expressing the continuity of the coordinates, velocity and acceleration at all
the points $s_{n},$ that is%
\begin{align*}
f^{\mu}(\tau)\text{\ }  &  \text{=}m\text{ }a_{s_{n}}^{\mu}(\tau)-\kappa\text{
}\frac{da_{s_{n}}^{\mu}(\tau)}{d\tau}-\kappa\text{ }a_{s_{n}}^{\nu}%
(\tau)a_{s_{n}\nu}(\tau)\text{ }u_{s_{n}}^{\mu}(\tau)\text{,}\\
\tau &  \in(s_{n},q_{n}),\\
x_{s_{n}}^{\mu}(s_{n}^{+})  &  =x_{q_{n}}^{\mu}(s_{n}^{-}),\\
u_{s_{n}}^{\mu}(s_{n}^{+})  &  =u_{q_{n}}^{\mu}(s_{n}^{-}),\\
a_{s_{n}}^{\mu}(s_{n}^{+})  &  =0.
\end{align*}

In the interior points of the intervals $(q_{n},$ $s_{n+1})$, since the
accelerations are vanishing the motion will be assumed to be uniform ones. The
coordinates and velocities, indicated with the subindex $q_{n},$ take the
expressions
\begin{align*}
x_{q_{n}}^{\mu}(\tau)  &  =\sum_{n=1}\theta(\tau-q_{n})\theta(s_{n+1}%
-\tau)(x_{n}^{\mu}(q_{n})+u_{n}^{\mu}(q_{n})(\tau-q_{n})),\\
u_{q_{n}}(\tau)  &  =\sum_{n=1}\theta^{(+)}(\tau-q_{n})\theta^{(-)}%
(s_{n+1}-\tau)u_{n}^{\mu}(q_{n}),\\
a_{q_{n}}^{\mu}(\tau)  &  =0.
\end{align*}

The being constructed solution for the coordinates in all the proper time
axis is written as the sum of the two expressions defined above. The
coordinate and their derivatives are continuous at the values of the proper
times thanks to their definition in the above equations.
\begin{align*}
x^{\mu}(\tau)  &  =\sum_{n=1}(x_{q_{n}}^{\mu}(\tau)\theta(\tau-q_{n}%
)\theta(s_{n+1}-\tau)+\\
&  x_{s_{n}}^{\mu}(\tau)\theta(\tau-s_{n})\theta(q_{n}-\tau)).
\end{align*}

It can be noted that the ALD equations are satisfied almost everywhere,
possibly except at the discrete set of points $s_{n}$ $q_{n}.$ Further, since
the coordinates and velocities are defined in a continuous way at those
points, but also the acceleration vanishes at the right of the points $s_{n}$
(due to the chosen initial conditions after the force is connected) the new
added term vanishes at these points and the proposed equation of motion is
also satisfied there. \ Thus, the equations could be not be obeyed only at the
ending points of the intervals $q_{n}$. However, although at those points
there is a discontinuity in the acceleration, the evaluation of the proposed
solution in the modified ALD equations in the sense of the generalized
functions leads to%

\begin{align*}
&  -f^{\mu}(\tau)\text{\ }+m\text{ }a^{\mu}(\tau)-\kappa\text{ }\frac{da^{\mu
}(\tau)}{d\tau}-m\text{ }a^{\nu}(\tau)a_{\nu}(\tau)\text{ }u^{\mu}(\tau)\text{
}+\kappa\text{ }a^{\mu}(\tau)\text{ }\sigma(\tau)\\
&  =\ \kappa\sum_{n=1}a_{s_{n}}^{\mu}(q_{n})\delta(\tau-q_{n})-\kappa
\sum_{n=1}a_{s_{n}}^{\mu}(q_{n})\delta^{(-)}(\tau-q_{n})\\
&  =0.
\end{align*}

\ Therefore the defined function satisfies the modified equations of motion.
The general set of equations in terms of the coordinates can be written in the
form%
\begin{align*}
f^{\mu}(\tau)\  &  =m\frac{d^{2}x^{\mu}(\tau)}{d\tau^{2}}-\kappa\frac
{d^{3}x^{\mu}(\tau)}{d\tau^{3}}-\kappa\frac{d^{2}x^{\nu}(\tau)}{d\tau^{2}%
}\frac{d^{2}x_{\nu}(\tau)}{d\tau^{2}}\frac{dx^{\mu}(\tau)}{d\tau}+\kappa
\frac{d^{2}x^{\mu}(\tau)}{d\tau^{2}}\sigma(\tau),\\
x^{\mu}(\tau_{0})\text{ }  &  \text{=\ }x_{0}^{\mu},\\
\frac{dx^{\mu}(\tau_{0})}{d\tau}  &  =u^{\mu}(\tau_{0})\text{ =\ }u_{0}^{\mu
},\\
\frac{d^{2}x^{\mu}(\tau_{0})}{d\tau^{2}}  &  =a^{\mu}(\tau_{0})\text{
=\ }0,
\end{align*}
in which initial conditions were fixed at given point internal to the intervals of vanishing
acceleration. \ \

At this point it seems helpful to recall that the sudden vanishing of the acceleration at the points
 $q_n$  was motivated by the Lienard-Wiechert field solution for a moving charge. Just at the moment
 at which a particle acquires a uniform motion, the field around it starts to become the Coulomb solution
 for a uniformly moving charge. However, on the contrary, if a charge is initially in uniform motion, and a
  force is suddenly applied, the radiation reaction, which is accepted to be defined by the time derivative of the
   acceleration, would be infinite, if the particle is assumed  to instantly get a finite value for the acceleration.
    Thus, the vanishing value of the acceleration fixed here  just after the instant of kind $s_n$ seems also  a natural
    physical choice.

Let us present here the solution for the simple case of the one dimensional
motion under \ a constant force which acts during a finite proper time
interval. The analysis of this problem led Dirac, as discussed in his classic
work on the theme, to the difficulties of the ALD equations in connection with
the runaway solutions. Consider the particle at the beginning being at the
origin of coordinates and that it moves with four velocity $u_{0}^{\mu}$ , at
that starting instant, taken as the origin of the proper time $\tau=0$. \ If
the motion is in the $x^{1}$ direction, the four-force and four-velocity have
the expressions
\begin{align*}
u^{\mu}(\tau)\  &  =\gamma(1,v,0,0),\\
\gamma &  =\frac{1}{\sqrt{1-v^{2}}},\\
f^{\mu}(\tau)\  &  =(F\text{ }v,F,0,0),\\
u^{\mu}(\tau)f_{\mu}(\tau)  &  =0,\\
u^{\mu}(\tau)u_{\mu}(\tau)  &  =1.
\end{align*}

The modified equations takes the explicit form
\begin{align*}
F\text{ }v\ \theta(\tau)\theta(T-\tau)  &  =m\frac{d^{2}x^{0}(\tau)}{d\tau
^{2}}-\kappa\frac{d^{3}x^{0}(\tau)}{d\tau^{3}}-\kappa(\frac{d^{2}x^{0}(\tau
)}{d\tau^{2}}\frac{d^{2}x_{0}(\tau)}{d\tau^{2}}-\frac{d^{2}x^{1}(\tau)}%
{d\tau^{2}}\frac{d^{2}x^{1}(\tau)}{d\tau^{2}})\frac{dx^{0}(\tau)}{d\tau
}+\kappa\frac{d^{2}x^{0}(\tau)}{d\tau^{2}}\sigma(\tau),\\
F\text{ }\ \theta(\tau)\theta(T-\tau)  &  =m\frac{d^{2}x^{1}(\tau)}{d\tau^{2}%
}-\kappa\frac{d^{3}x^{1}(\tau)}{d\tau^{3}}-\kappa(\frac{d^{2}x^{0}(\tau
)}{d\tau^{2}}\frac{d^{2}x_{0}(\tau)}{d\tau^{2}}-\frac{d^{2}x^{1}(\tau)}%
{d\tau^{2}}\frac{d^{2}x^{1}(\tau)}{d\tau^{2}})\frac{dx^{1}(\tau)}{d\tau
}+\kappa\frac{d^{2}x^{1}(\tau)}{d\tau^{2}}\sigma(\tau),\\
x^{0}(0)  &  =x^{0}(0)=0,\\
\frac{dx^{0}(0)}{d\tau}\text{ }  &  \text{=\ }\gamma,\text{ \ \ \ }%
\frac{dx^{1}(0)}{d\tau}=\gamma\text{ }v,\\
\frac{d^{2}x^{0}(0)}{d\tau^{2}}  &  =0\text{ },\text{ \ }\frac{d^{2}x^{1}%
(0)}{d\tau^{2}}=0.\text{ }%
\end{align*}

The parameters and initial conditions fixed were $\kappa=1, m=10,
f =10, v  =0.5$ and the  proper time interval in which the force acts was  selected as $T=0.1$.
The solution for the acceleration $\frac{d^{2}x^{1}(\tau)}{d\tau^{2}}$ of
the equations is illustrated in figure 1. \ The basic point to note is that
the solution in the interval of proper time in which the force acts, is the
same as the one for the ALD equation. However, when the force is removed the
acceleration instantly vanishes. Thus, the only function of the added term in
the ALD equation has been simply to facilitate the system to suddenly attains
its uniform motion state, also satisfying the ALD equation. In the case of the
ALD equation, discussed by Dirac in the reference \cite{Dirac}, the
acceleration, after the force is eliminated, is not vanishing due to the
continuity of the acceleration, and more troublesome, it tends to grow without
limit.  The application of this particular solution to developing a discretization of a general
 motion, searching for avoiding the  numerical development of instabilities leading to  runaway solutions
will be  considered in extensions of this work.
\begin{figure}[h]
\begin{center}
\hspace*{-0.4cm} \includegraphics[width=7.5cm]{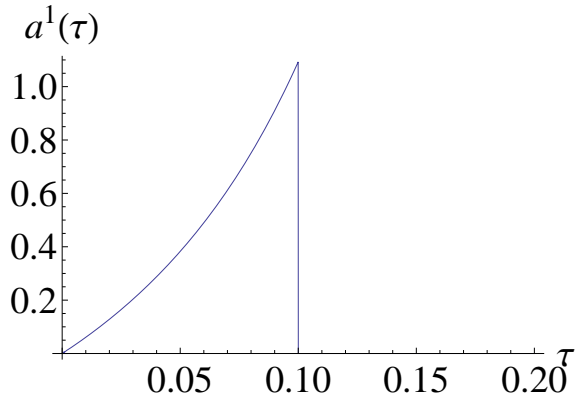}
\end{center}
\caption{ The figure shows the solution of the  equations under study for the important case of a force which is suddenly applied 
remains constant during a time interval $T=0.1$  and afterwards is also suddenly  removed. The parameters of the equation fixed 
were $\kappa=1, m=10, f =10, v  =0.5$    }%
\label{fig1}%
\end{figure}\

In this sense, the present discussion seems to be close connected with
the analysis done in Ref. \cite{Yighjian}, which underlines, that in the zones
of the evolution in which the forces have  a non analytic behavior, the ALD
equations can not be consistently derived and alternative equations should be
obeyed. \ Therefore, the added term could represent an allowed modification of
the dynamics, being able to solve the basic difficulties of the ALD
equations, and which is determined by the physically plausible condition that the
 self-force should vanish instantaneously after the sudden (non-analytical) removal of the external force.
 This reasonable nature of this last condition is strongly supported  by the fact that exactly after
  the instant of removal of the force,  the wave-like propagation of the field, suggests that it
   starts to become exactly given by a boosted Coulomb field in sufficiently small neighborhood of the particle.

The solution of equations for more complex situations will be explored in the
extension of the work. \ Of particular interest seems its application in
improving numerical calculations of the solutions of the ALD equations, which
had been systematically plagued with instabilities due to the existence of the
nonphysical runaway solutions \cite{Yighjian}. \ One specific problem  will be
the application to the one dimensional motion of a particle under the  action
of a repulsive Coulomb potential. In this task the runaway instabilities
appear and their elimination had not been attained \cite{voigt}.\\

The authors would like to deeply acknowledge the helpful comments on the work received from  Prof. Danilo Villarroel (Chile) and  Prof. Luis Carlos Bassalo Crispino (UFPA).
One of the authors (A.C) strongly acknowledges the support received from the Coordenac\~{a}o de Aperfeicoamento
de Pessoal de N\'{\i}vel Superior (CAPES) of Brazil and the Postgraduation
Programme in Physics (PPGF) of the Federal University of Par\'{a} at
Bel\'{e}m, Par\'{a} (Brazil), in which this work was done, in the context of a
CAPES External Visiting Professor Fellowship. The support also received by
(A.C.) from the Caribbean Network on Quantum Mechanics, Particles and Fields
(Net-35) of the ICTP Office of External Activities (OEA), the "Proyecto
Nacional de Ciencias B\'{a}sicas"(PNCB) of CITMA, Cuba is also very much acknowledged.

\end{document}